\documentclass[
superscriptaddress,
reprint,
 amsmath,amssymb,
 aps,
]{revtex4-2}

\usepackage{upgreek}
\usepackage{graphicx}
\graphicspath{{figures/}}
\usepackage{dcolumn}
\usepackage{bm}
\usepackage{hyperref}
\usepackage[mathlines]{lineno}
\usepackage{ulem}
\usepackage{multirow}
\usepackage{booktabs}
\usepackage{caption}
\usepackage{subcaption}
\usepackage{siunitx}
\usepackage{xcolor,colortbl}
\usepackage{array}
\usepackage{tabularx}
\usepackage{footmisc}

\begin{document}


\title{Lifetimes of excited states in Rh$^-$}

\author{J. Karls}
\affiliation{%
 Department of Physics, University of Gothenburg, SE-412 96 Gothenburg, Sweden
}%
\author{J. Grumer}
\affiliation{
Department of Physicsand Astronomy, Theoretical Astrophysics, Uppsala University, Box 516, SE-75120 Uppsala, Sweden
}%
\author{S. Schiffmann}
\affiliation{
Division of Mathematical Physics, Department of Physics, Lund University, SE-22100 Lund, Sweden
}%
\affiliation{
 Spectroscopy, Quantum Chemistry and Atmospheric Remote Sensing (SQUARES), CP160/09, Université libre de Bruxelles (ULB), B-1050 Bruxelles, Belgium
}%
\author{N. D. Gibson}
\affiliation{%
 Department of Physics and Astronomy, Denison University, Granville, Ohio 43023, USA
}%
\author{M. Ji}
\affiliation{%
 Department of Physics, Stockholm University, AlbaNova, SE-106 91 Stockholm, Sweden
}%
\author{M. K. Kristiansson}
\affiliation{%
 Department of Physics, Stockholm University, AlbaNova, SE-106 91 Stockholm, Sweden
}%
\author{D. Leimbach}
\affiliation{%
 Department of Physics, University of Gothenburg, SE-412 96 Gothenburg, Sweden
}%
\author{J. E. Navarro Navarrete}
\affiliation{%
 Department of Physics, Stockholm University, AlbaNova, SE-106 91 Stockholm, Sweden
}%
\author{Y. Pe\~{n}a Rodr\'{i}guez}
\affiliation{%
 Department of Physics, University of Gothenburg, SE-412 96 Gothenburg, Sweden
}%
\author{R. Ponce}
\affiliation{%
 Department of Physics and Astronomy, Denison University, Granville, Ohio 43023, USA
}%
\author{A. Ringvall-Moberg}
\affiliation{%
 Department of Physics, University of Gothenburg, SE-412 96 Gothenburg, Sweden
}%
\author{H. T. Schmidt}
\affiliation{%
 Department of Physics, Stockholm University, AlbaNova, SE-106 91 Stockholm, Sweden
}%
\author{S. E. Spielman}
\affiliation{%
 Department of Physics and Astronomy, Denison University, Granville, Ohio 43023, USA
}%
\author{C. W. Walter}
\affiliation{%
 Department of Physics and Astronomy, Denison University, Granville, Ohio 43023, USA
}%
\author{T. Brage}
\affiliation{
Division of Mathematical Physics, Department of Physics, Lund University, SE-22100 Lund, Sweden
}%
\author{D. Hanstorp}
\affiliation{%
 Department of Physics, University of Gothenburg, SE-412 96 Gothenburg, Sweden
}%

\date{\today}

\begin{abstract}
The radiative decay of excited states of the negative ion of rhodium, Rh$^-$, has been investigated experimentally and theoretically. The experiments were conducted at the Double ElectroStatic Ion Ring Experiment (DESIREE) facility at Stockholm University using selective photodetachment from a stored ion beam to monitor the time evolution of the excited state populations. The lifetimes of the Rh$^-$ $^3F_{3}$ and $^3F_{2}$ fine structure levels were measured to be 3.2(6)~s and 21(4)~s, respectively. An additional, previously unreported, higher-lying bound state of mixed $^1D_2+^3P_2+(4d^95s)^1D_2+^3F_2$ composition was observed and found to have a lifetime of 10.9(8)s. The binding energy of this state was determined to be in the interval $0.1584(2) $ eV $ < E_b < 0.2669(2)$ eV, using laser photodetachment threshold (LPT) spectroscopy. An autodetaching state with a lifetime of 480(10)$~\upmu$s was also observed. Theoretical calculations of the excited-state compositions, energies, and magnetic-dipole transition lifetimes were performed using the multiconfiguration Dirac-Hartree-Fock and relativistic configuration interaction methods. The calculated lifetimes of the $^3F_{3}$ and $^3F_{2}$ fine structure levels are in excellent agreement with the measured values.  The present study should provide valuable insights into electron correlation effects in negative ions and forbidden radiative transitions.
\end{abstract}

\maketitle


\section{\label{sec:introduction}Introduction}
Negative ions are very different from positive ions or neutral atoms in the sense that there is no long-range Coulomb potential. Instead, the valence electron in a negative ion is bound by a short-range induced dipole potential, making the electron correlation effect very important. Thus, the independent particle model, breaks down for negative ions. This also means that negative ions only have a few, if any, bound excited states and most transitions within negative ions are optically forbidden. The only elements known to have excited states with opposite parity from the ground state are Os$^-$, Ce$^-$, La$^-$, Th$^-$ and U$^-$ \cite{Bilodeau2000ExperimentalIon,Walter2011ExperimentalCe-,Kellerbauer2014MeasurementOs-,Walter2014CandidateLa,Tang2019CandidateTh-,Tang2021ElectronAnion}. In all other known cases, electric dipole (E1) transitions are not possible and the decay modes from excited states to the ground state will be magnetic dipole (M1) or electric quadrupole (E2) transitions. An interesting example of the latter is the very recent observation of a laser-induced E2 transition in bismuth \cite{Walter2021}. These higher order transitions will typically result in considerably longer radiative lifetimes, and the need for long ion beam storage times hence becomes crucial. 

Studies of the lifetimes of excited states of negative ions provide valuable opportunities to gain insight into electron correlation effects. The first experiments measuring lifetimes in negative ions were performed on a metastable autodetaching state in He$^-$  \cite{Blau1970LifetimesIon}. Since then, the radiative lifetimes of excited states in several negative ions have been studied \cite{Andersson2006RadiativeIons,Ellmann2004RadiativeTe-,Backstrom2015Storing-}. The ion storage rings available today, e.g. DESIREE in Stockholm, where negative ions can be stored for up to an hour \cite{Backstrom2015Storing-}, allow for very long lifetimes to be measured. Recent investigated elements at DESIREE are Ir$^-$, Bi$^-$ \cite{Kristiansson2021,Kristiansson2022MeasurementBi-} and at the Cryogenic Storage Ring (CSR) at the Max Planck Institute for Nuclear Physics in Heidelberg lifetimes of metastable states in Si$^-$ have been measured \cite{Mull2021MetastableRing}. 

\begin{figure}
    \centering
    \includegraphics[trim={0cm 0cm 0cm 0cm},clip,width = 0.5\columnwidth]{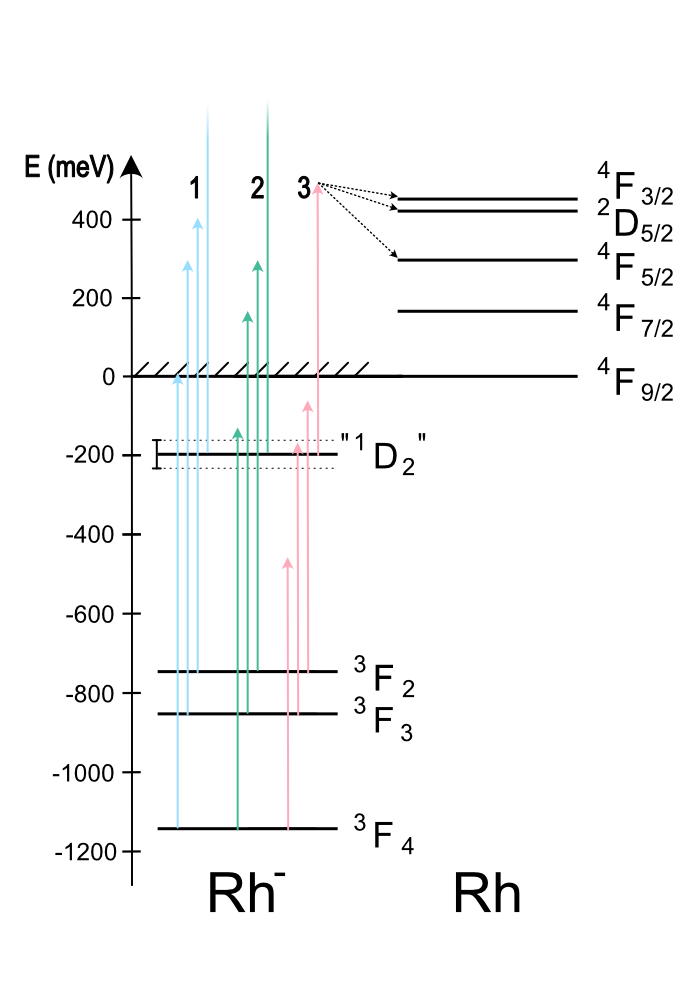}
    \caption{Energy level diagram of Rh$^-$. Arrow set 1 (blue) corresponds to a laser wavelength of 1000 nm, arrow set 2 (green) corresponds to a laser wavelength of 1100 nm and arrow set 3 (red) corresponds to a laser wavelength of 1800-2000 nm. This energy level diagram shows the energy levels as interpreted by the results in this paper. For clarity, this is presented upfront and will be explained in sections \ref{sec:results} and \ref{sec:discussion}.}
    \label{fig:energylevels}
\end{figure}

In this work, we investigate excited states of the negative ion of rhodium, Rh$^-$, both experimentally and theoretically. Rhodium (atomic number 45) is an element in the platinum group of transition metals and has a low natural abundance. $^{103}$Rh is the only naturally occurring stable isotope. The atomic ground state of Rh is [Kr]$4d^85s$ $^4F_{9/2}$, whereas the ground state of Rh$^-$ is [Kr]$4d^85s^2$ $^3F_{4}$, as shown in the energy level diagram in Fig. \ref{fig:energylevels}.  The electron affinity of Rh, defined as the binding energy of the $^{3}F_{4}$ anionic ground state, has previously been measured to be $1.142~89(20)$~eV \cite{Scheer1998LaserPd}. Further, two bound excited fine structure states, $^3F_{3}$ and $^3F_{2}$, have been observed, where the $J=3-4$ and $2-4$ splittings were measured to be 0.294(8)~eV and 0.418(8)~eV, respectively~\cite{Feigerle1980BindingIons}. 

In this work, the experimental measurements of radiative lifetimes of the Rh$^-$ $^3F_{3}$ and $^3F_{2}$ fine structure levels were performed at the DESIREE facility, using laser photodetachment studies of the different energy states. In addition, another, previously unobserved, state of Rh$^-$ with a weaker binding energy was found. The binding energy of this state was measured using LPT spectroscopy at the Gothenburg University Negative Ion Laser Laboratory (GUNILLA) in Gothenburg and is presented in the energy level diagram of Rh$^-$ in Fig. \ref{fig:energylevels}. Further, an unbound state near just above the detachment limit was predicted theoretically. The first three states in Rh$^-$ are labelled $^3F_J$. For convenience, the bound mixed state is labelled ``$^1D_2$" and the unbound mixed  state is labelled ``$^3P_2$" from the leading terms. The detailed $LS$-compositions are presented in table \ref{tab:LScomp}.

The theoretical calculations of excited state energies and lifetimes in this work have been conducted using multiconfiguration Dirac-Hartree-Fock (MCDHF) and relativistic interaction configuration (RCI) methods \cite{Fischer2016AdvancedFunctions} as implemented in the \textsc{Grasp2018} package \cite{Fischer2019GRASP2018-APackage}.  
Combining theoretical and experimental studies is essential to understand the impact of electron correlation on the binding energies and lifetimes of excited states in negative ions. 

\section{\label{sec:experimentalmethods}Experimental methods}
\begin{figure*}
    \centering
     \begin{subfigure}[b]{0.7\textwidth}
         \centering
    \includegraphics[trim={0cm 5cm 0cm 4.5cm},clip,width = 0.8\columnwidth]{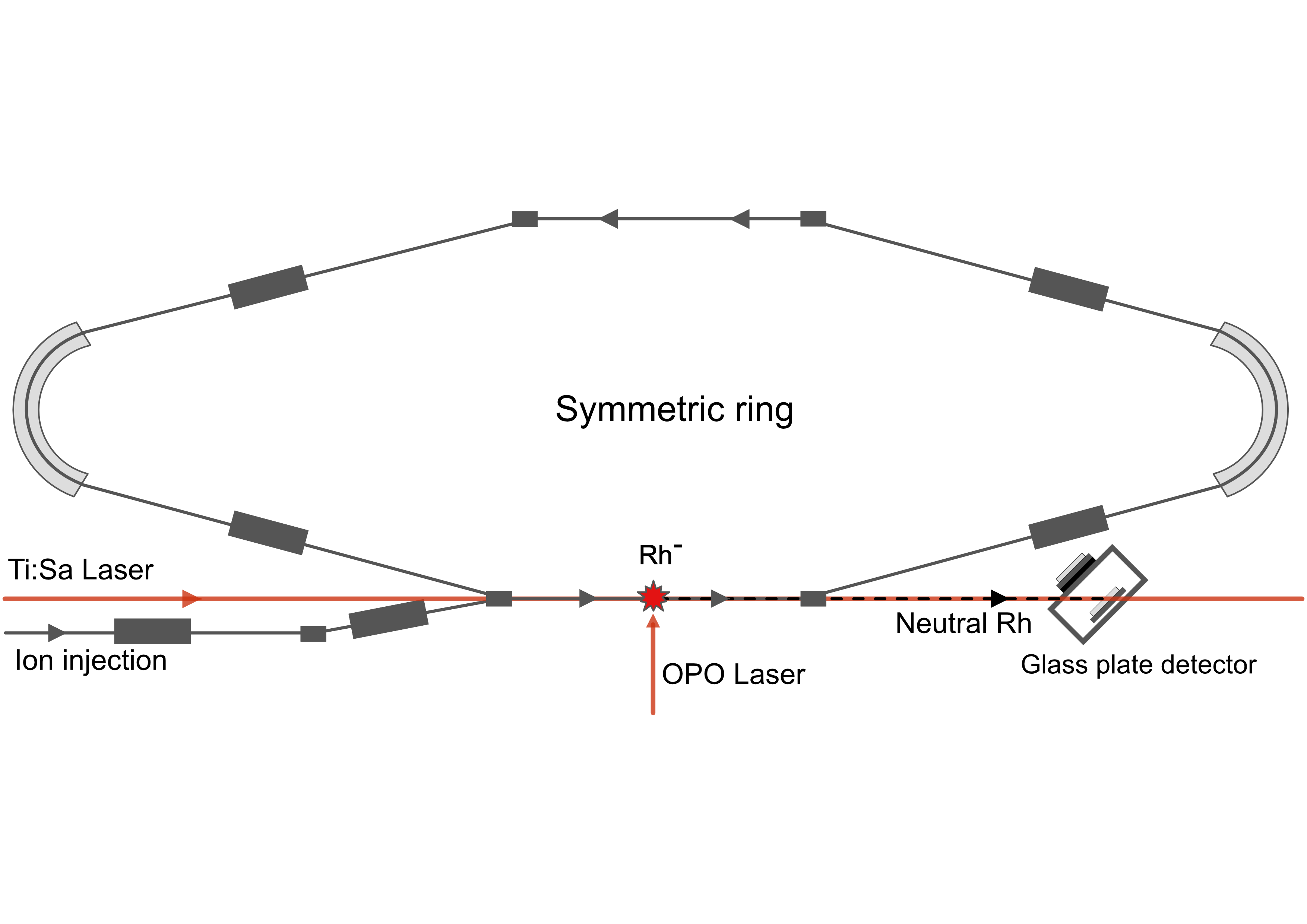}
    \caption{Schematic of the symmetric storage ring at DESIREE.}
    \label{fig:DESIREE}
    \end{subfigure}
     \begin{subfigure}[b]{0.29\textwidth}
         \centering
    \includegraphics[trim={0cm 0cm 0cm 0cm},clip,width = \columnwidth]{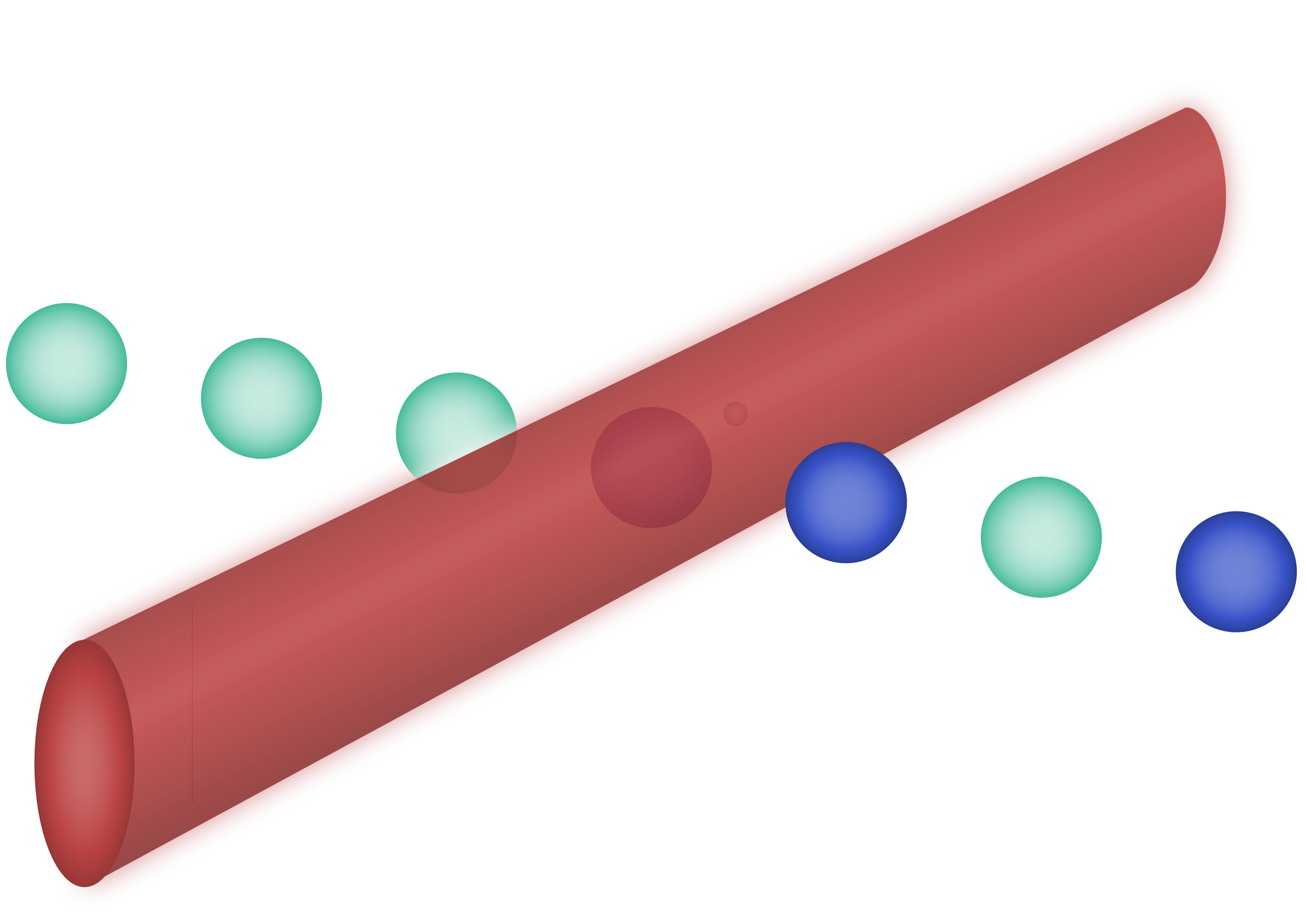}
    \caption{Ion-laser interaction region.}
    \label{fig:interaction_reg}
    \end{subfigure}
    \caption{In the setup, two different laser beams are used. An Optical Parametric Oscillator (OPO) operating in a pulsed mode and a continuous wave Titanium:Sapphire (Ti:Sa) laser. The OPO  intersects the ion beam at a \SI{90}{\degree} angle, whereas the Ti:Sa exhibits a collinear alignment relative to the ion beam.}
    \label{fig:setup}
\end{figure*}

The experiment was carried out at one of the two storage rings at the cryogenic DESIREE (Double ElectroStatic Ion Ring ExpEriment) facility, as illustrated in Fig. \ref{fig:DESIREE}. The storage rings are kept at a temperature of 13.7 K and a pressure of $10^{-14}$ mbar. This reduces the residual gas collisions and allows for long ion beam storage times \cite{Backstrom2015Storing-}. The circumference of each storage ring is 8.7 m. The facility is described in detail in previous publications \cite{Thomas2011TheStudies,Schmidt2013FirstDESIREE,Backstrom2015Storing-}. 

Rh$^-$ ions are produced in a SNICS II cesium sputter source \cite{NationalElectrostaticsCorporationSourceII} and accelerated to an energy of 10 keV, which means the excited states are highly populated. The ions are then mass selected in a mass-separating magnet and guided through the system to the storage ring using ion beam optics. Before injection into the storage ring, the ion beam current is measured using a Faraday cup. In order to avoid ion-ion collisions, the ion beam current is reduced to a level where they do not have an impact on the measured lifetimes. The ion beam current is thus typically kept around 1 nA, but never above 4 nA for any measurement. 

In the storage ring, the ion beam can be overlapped with a laser beam in two straight sections. This can be done in either collinear or crossed-beam geometry. In this experiment, the ion beam was intersected at a \SI{90}{\degree} angle by an optical parametric oscillator (OPO) laser pumped by a diode-pumped nanosecond Q-switched Nd:YAG laser at a repetition rate of 1 kHz or another OPO system pumped by a Q-switched Nd:YAG laser at a repetition rate of 10 Hz. Both systems were used for studying the shorter-lived lower-lying $^3F_{3}$ state and the longer-lived $^3F_{2}$ state, whereas only the 1 kHz laser was used for studying the mixed bound state in order to obtain sufficient statistics. 
The storage time of the total ion beam in the storage ring was measured by detaching all states in the negative ion with a continuous-wave Matisse titanium:sapphire (Ti:Sa) laser in a collinear geometry. The laser was chopped with a shutter opened for 0.5 s and shut for 9.5 s in order to measure the non-laser produced background due to collisional stripping and dark counts. 

After the interaction region, the photodetached neutral atoms continue along a straight line until they enter a glass plate Resistive Anode Encoder (RAES) detector, which consists of a gold and titanium coated glass plate. When the atoms impinge on its surface, secondary electrons are emitted and accelerated by an electric field into three stacked microchannel plates (MCPs) where they are amplified and finally detected by an anode.

Intersecting the ion beam with a laser beam at different photon energies allows for selective photodetachment of the separate energy levels. The lifetimes of the excited states were then extracted by making a fit of exponential decay to the data. 

In addition to the lifetime measurements, laser photodetachment threshold (LPT) spectroscopy measurements were performed at the GUNILLA facility in Gothenburg \cite{Hanstorp1995AnExperiments,Diehl2011IonApparatus}. A photodetachment threshold was observed when systematically scanning the photon energy of a 10 Hz OPO laser overlapped with an Rh$^-$ ion beam in a  co- and counter-propagating geometry.
This threshold and the energy for the observed transition was compared to the known energy levels in the neutral Rh atom to deduce the binding energy for the mixed bound state.

\section{\label{sec:theory}Theoretical methods}
We performed atomic structure calculations using the multiconfiguration Dirac-Hartree-Fock (MCDHF) and relativistic interaction configuration (RCI) methods (for a review see~\cite{Fischer2016AdvancedFunctions}) as implemented in the \textsc{Grasp2018} package~\cite{Fischer2019GRASP2018-APackage}.  Electron correlation is in these methods accounted for by expanding the atomic state function over a set of basis functions, labelled Configuration State Functions (CSFs). These are generated by allowing for electron substitutions from a set of the most important configurations, namely, the multireference (MR) set. The size of the expansion is important, but it is essential to include the correct CSFs in the MR-set, to capture the most important correlation. One example is, that often one includes only single and double excitations of electrons from the MR-set, but the most important triple excitations could be included by a correct selection of the MR-set. Therefore, this work started with exploratory calculations, where it was realized that it is important to include the $4d^9 5s$ configuration as part of the MR in addition to the targeted $4d^85s^2$ configuration.  The correlation model was chosen following recent calculations carried out using the same MCDHF/RCI method for Ir$^-$ ($Z=77$) \cite{Kristiansson2021}, which is homologous to Rh$^-$ and exhibits a similar atomic structure. In line with the observations made in Ir$^-$, electron correlation was included by allowing for single and double (SD) substitutions from the $4d$ and $5s$ valence orbitals of the MR configurations, capturing most of the valence-valence (VV) correlation. In subsequent RCI calculations, the closed shells of the $4s^24p^6$ core were open to substitutions, through which important core-valence (CV) correlation contributions are included. The two correlation models are referred to as VV and VV+CV, respectively. In both correlation models the Breit interaction and the leading quantum electrodynamics (QED) corrections (self-energy and vacuum polarization) are included.

\section{\label{sec:results}Results}
The structures of the group VIII elements, including Rh and Ir, are known to be sensitive to electron correlation. This is apparent, for example, from the change in the ground configurations of the neutral atom between the homologous systems Rh ($4d^8 5s$) and Ir ($5d^7 6s^2$). However, the ground configurations of the anions have the homologous structure, $nd^8(n+1)s^2$, with $n=4$ for Rh$^-$ and $n=5$ for Ir$^-$. Just as for Ir$^-$, four states of Rh$^-$ are predicted to be bound according to their excitation energies displayed in Table~\ref{tab:LScomp}. That is, one more than established prior to this work. 

In terms of radiative properties, we computed both transition rates and radiative lifetimes. Since all bound states have the same parity, electric dipole (E1) transitions are forbidden and the bound excited states can only decay through higher-order transitions with the dominant being the magnetic dipole (M1) and electric quadrupole (E2) contributions. However, because the M1 transition rates were found to be significantly larger than the E2 transitions rates (at least by 2 orders of magnitudes), the radiative lifetimes were approximated to
\begin{equation}
    \tau(\Gamma J) = \frac{1}{\sum_{J'} A_{\rm M1}(\Gamma J\to \Gamma 'J')} \ ,
\end{equation}
where $A_{\rm M1}(\Gamma J\to \Gamma' J')$ is the rate of the M1 transition between the upper and lower atomic levels $\Gamma J$ and $\Gamma' J'$, respectively. Here $\Gamma$ represents all information such as other quantum numbers or configurations to uniquely define the level of interest.
The M1 rates are often considered more reliable, particularly since they are also independent of the wave functions, to first order~\citep{Grant2006,Cowan1981}.

\subsection{\label{sec:storage} Beam lifetime}
\begin{figure*}
    \centering
     \begin{subfigure}[b]{0.3\textwidth}
         \centering
         \includegraphics[trim={0cm 0cm 0cm 0cm},clip,width=\textwidth]{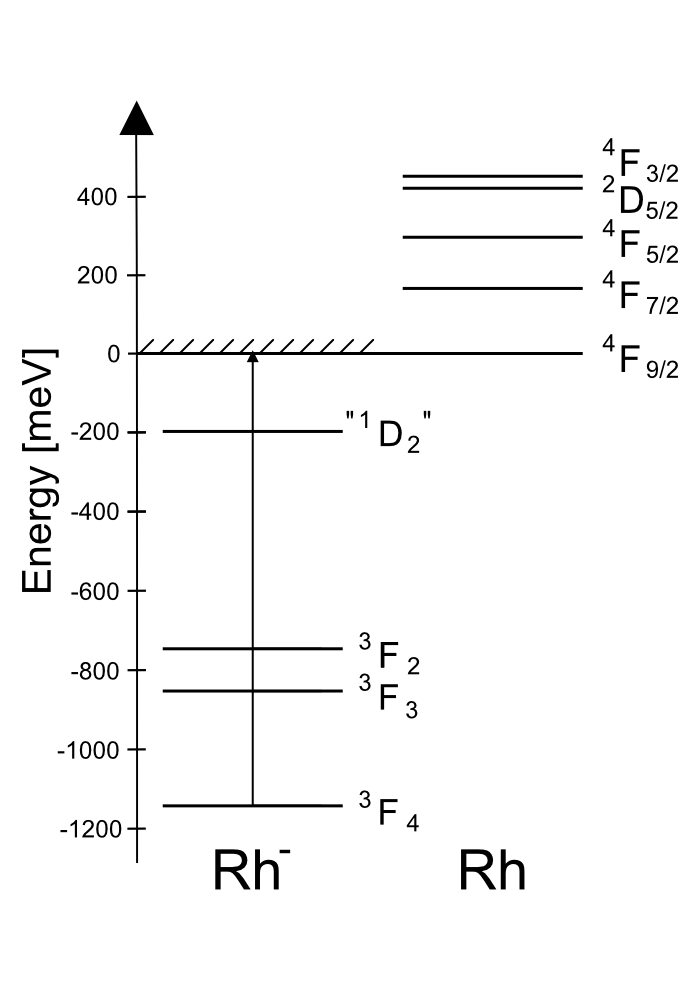}
     \end{subfigure}
        \hfill
    \begin{subfigure}[b]{0.6\textwidth}
        \includegraphics[width = \columnwidth]{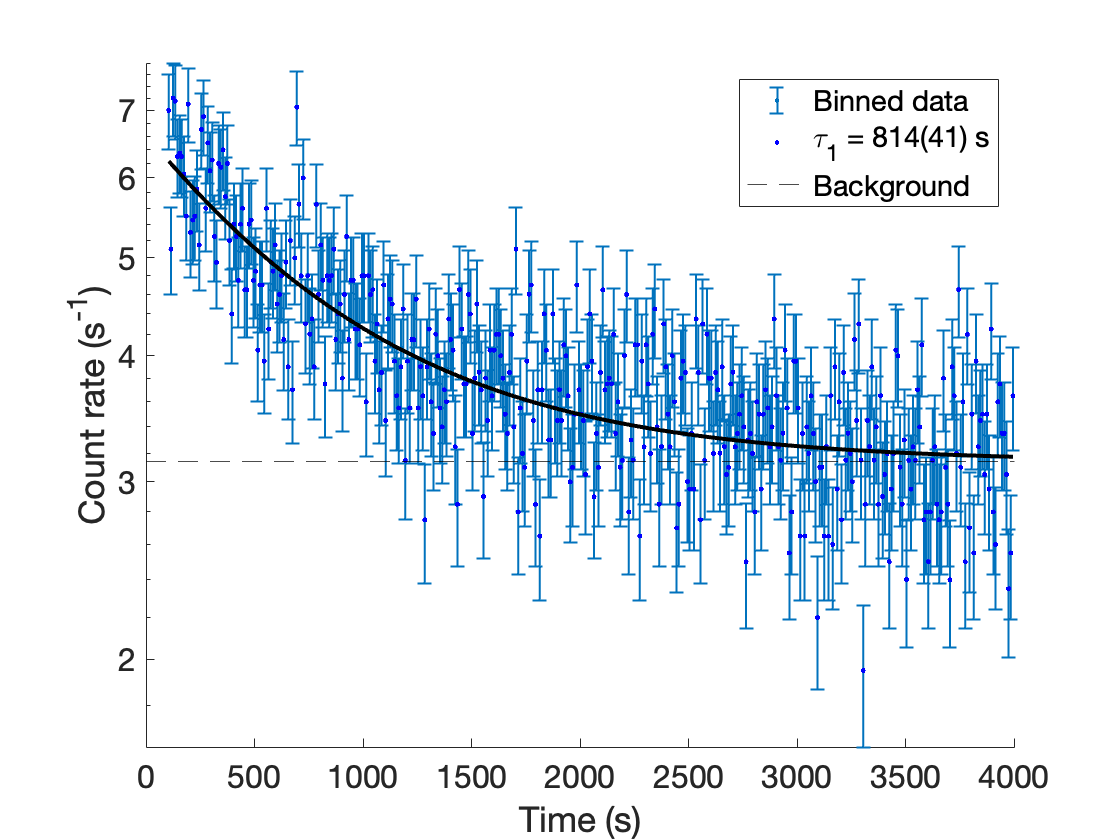}
    \end{subfigure}
        \hfill
    \caption{Total ion beam storage time in the storage ring.}
    \label{fig:beamstoragetime}
\end{figure*}

The total storage time of the ion beam was determined by fitting an exponential decay to the data recorded during a 4000 s measurement cycle. This data and exponential fit are shown in Fig. \ref{fig:beamstoragetime}. The data was collected using a chopped continuous wave Ti:Sa laser in a collinear setup with a photon energy of 1.24 eV. This photon energy is well above the photodetachment threshold for the ground state in Rh$^-$, which means that all bound states in the ion are detached. The resulting ion beam storage time extracted from the exponential decay curve was 814(41) s, which gives us an upper limit for the lifetime measurements.

\subsection{\label{sec:3F3_3F2} The $^3F_3$ and $^3F_2$ states}
Looking at the $LS$-term composition of the bound states shown in Table~\ref{tab:LScomp}, we find that, within the VV correlation model, the three lower levels with $J=4,3,2$ are all dominated by their $^3\!F$ component with a weight well over $80\%$. 

Radiative lifetimes are shown in Table~\ref{tab:LScomp}. All bound levels have lifetimes of similar order of magnitudes, i.e., of a few seconds. Only the lowest $J=2$ stands out with its slightly larger lifetime of $\approx 20$~s. We observe small discrepancies in the computed lifetimes between the two correlation models. To reduce these differences, we can compute lifetimes adjusted to the experimental transition energy (see, e.g.~\cite{Su2019}). The resulting adjusted lifetimes are also displayed in Table~\ref{tab:LScomp}. As expected the discrepancies between the two models decrease and we derive the theoretical lifetimes $\tau=2.9(2)$~s and $\tau=20.2(4)$~s based on the VV correlation model for the two lowest excited levels for which experimental energies are available.

\begin{figure*}
    \centering
     \begin{subfigure}[b]{0.3\textwidth}
         \centering
         \includegraphics[trim={0cm 0cm 0cm 0cm},clip,width=\textwidth]{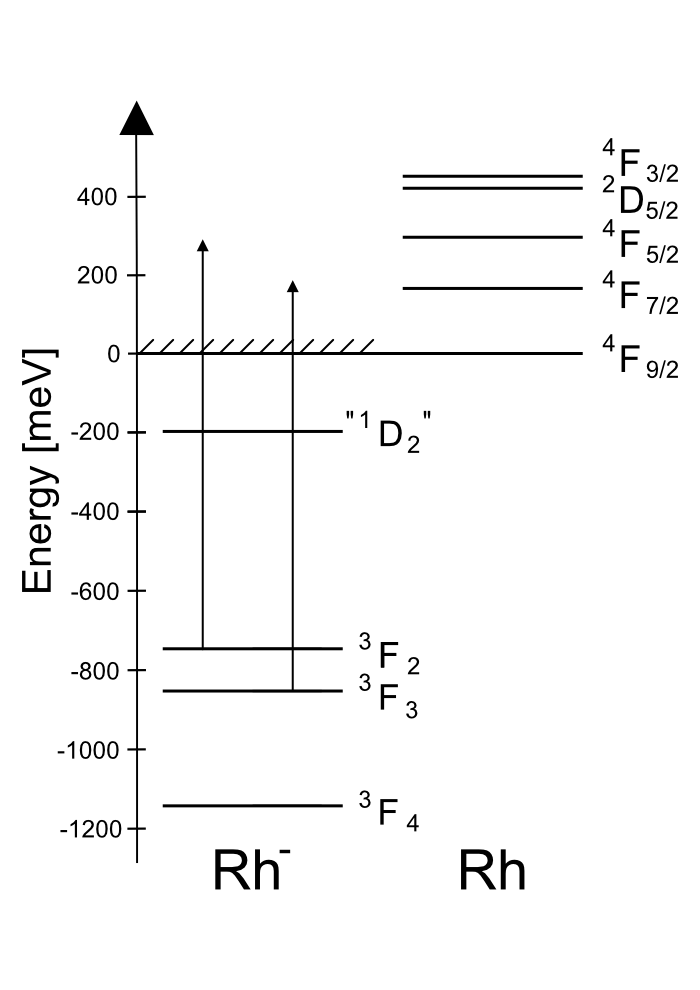}
     \end{subfigure}
        \hfill
    \begin{subfigure}[b]{0.6\textwidth}
        \includegraphics[width = \columnwidth]{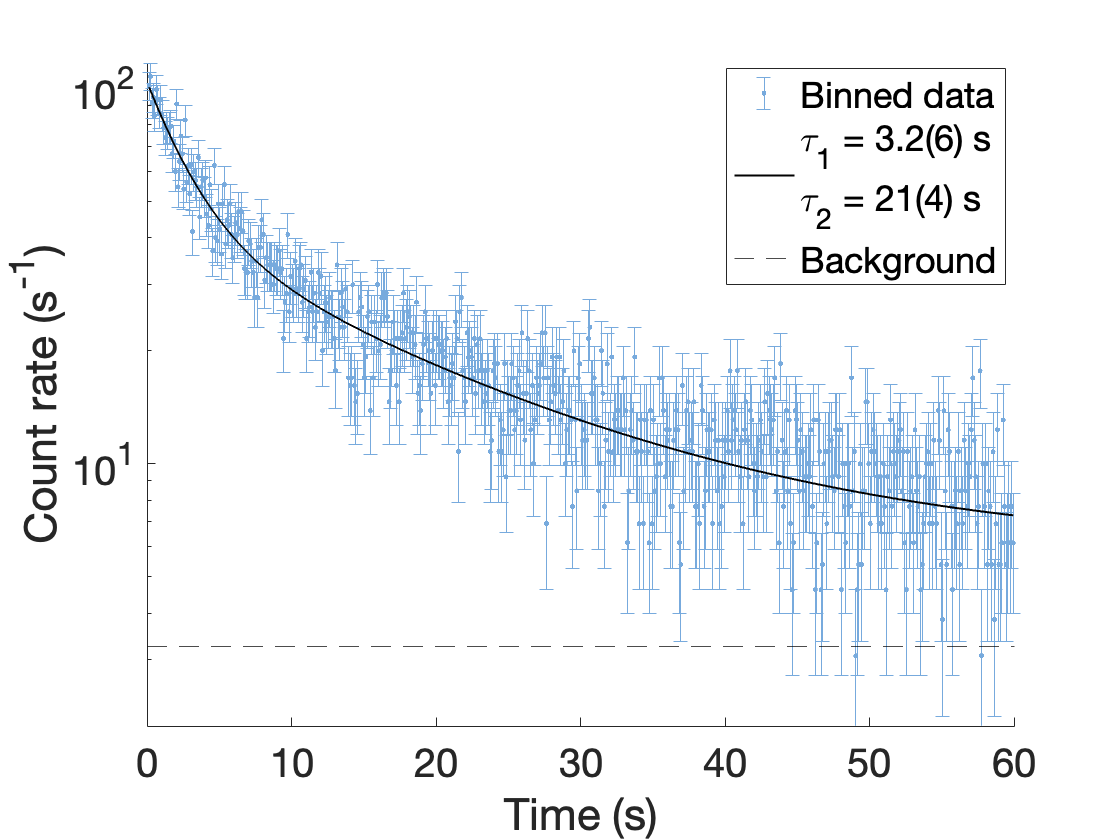}
    \end{subfigure}
        \hfill
    \caption{A typical lifetime measurement of the $^3F_3$ and $^3F_2$ states.}
    \label{fig:longandshort1100nm}
\end{figure*}

The energies of the lower lying bound excited states, $^3F_3$ and $^3F_2$, have been previously measured, and the energy splitting between $^3F_3$ and the $^3F_4$ ground state is 0.294(8) eV, whereas the energy splitting between $^3F_2$ and $^3F_4$ is 0.418(8) eV \cite{Feigerle1980BindingIons}. Due to the difficulty of resolving the two thresholds ($^3F_3 \rightarrow a^4F_{7/2}$ and $^3F_2 \rightarrow a^4F_{5/2}$), they were measured at the same photon energy. Making a double exponential decay fit results in two different lifetimes of 3.2(6) s and 21(4) s. The two lifetimes can then be separated using the theoretical calculations of the lifetimes, where the lifetime of the $^3F_2$ state is calculated to be one order of magnitude larger than that of the $^3F_3$ state.  Thus we can assign the shorter lifetime  of 3.2(6) s to the $^3F_3$ state and the longer lifetime of 21(4) s to the  $^3F_2$ state. The double exponential fit and data for these two states are shown in Fig. \ref{fig:longandshort1100nm}.

\subsection{\label{mixed} The highly mixed excited states}

The second $J=2$ level, is mixed with contributions from $^{1}\!D$ (57$\%$) and $^{3}\!P$ (27$\%$). For short, we will label this  ``$^{1}\!D_2$". Finally, a third $J=2$ level is also studied and is found to be highly mixed and unbound. We will label this state ``$^{3}\!P_2$", from the leading contribtuion. Its highly mixed $LS$-composition suggests a strong interaction with the ``$^{1}\!D_2$" state. According to the results of the calculations in the VV+CV model, also displayed in Table~\ref{tab:LScomp}, we observe that the three lower bound levels belonging to the $^{3}\!F$ fine-structure have similar $LS$-compositions compared to those computed in the VV correlation model calculations. Oppositely, the ``$^{1}\!D_2$" and ``$^{3}\!P_2$" states, with higher excitation energies, exhibit significant differences in their $LS$-compositions when comparing the two models, confirming the existence of a strong interaction between these two levels. The introduction of CV correlation increases the $^3F_4$ to $^3F_3$ fine structure splitting by 13~meV ($\sim$ 105~cm$^{-1}$), as expected for fine structure splittings~\citep{Fischer1997}. However, the theory--experiment discrepancy increases for the $^3F_3$ to $^3F_2$ fine structure splitting, probably due to the presence of the ``$^{1}\!D_2$" level. For the ab initio calculation of the lifetime of the ``$^{1}\!D_2$" state, it is thus reasonable to suggest $\tau=1.6(2)$~s taken from the more advanced VV+CV correlation model, where the theoretical uncertainty estimation is set to the difference between the lifetimes computed in the two different correlation models.

\begin{figure}
    \centering
    \includegraphics[width = 0.99\columnwidth]{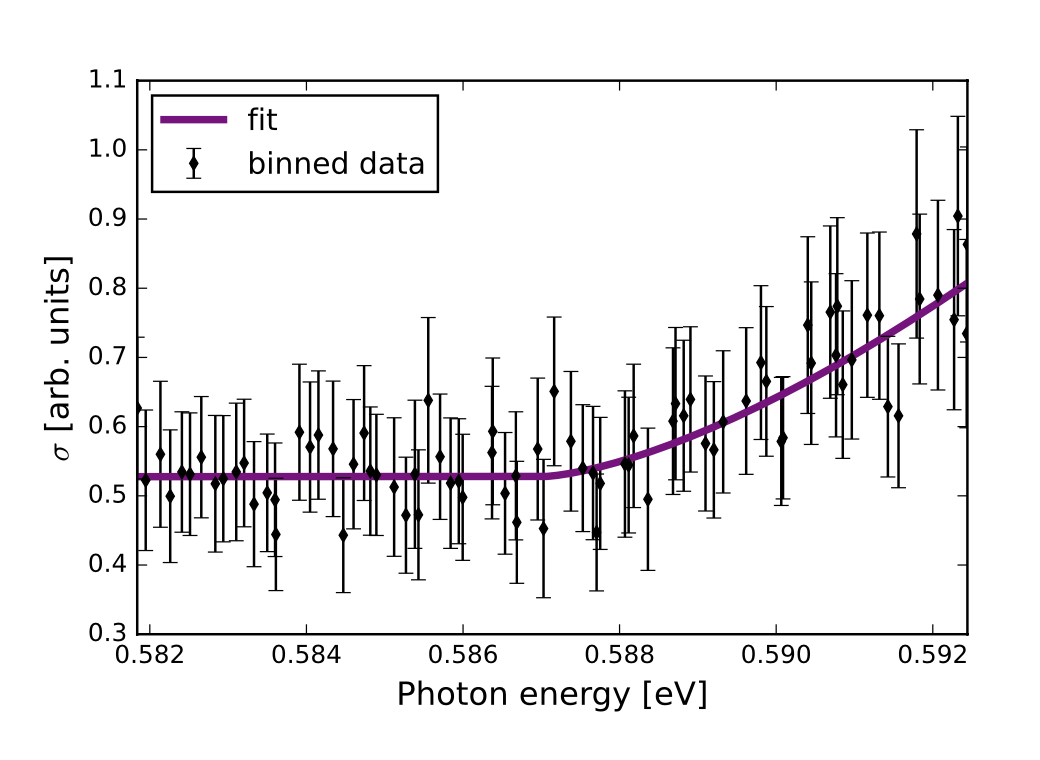}
    \caption{Scan of the $p$-wave threshold for the mixed bound state in Rh$^-$.}
    \label{fig:threshold}
\end{figure}

The ``$^1D_2$" state has not been observed previously, so the energy threshold for this state was investigated experimentally and a typical scan over the threshold is shown in Fig. \ref{fig:threshold}. 
After comparing the measured threshold energy of $0.5890(2)$ eV to the known energy levels in the neutral Rh atom, the possible observed transitions are presented in table \ref{tab:mixed_transition}. From GRASP calculations we can see that detachment of an $s$-electron in Rh$^-$ is much more likely than detachment of a $d$-electron. Thus the observed transition is likely to be to the $^3\!F_{3/2}$, $^3\!F_{5/2}$ or $^2\!D_{5/2}$ state in Rh I. Also, since no threshold was observed at higher photon energies, the observed threshold must correspond to one of these transitions with a larger energy gap. Otherwise we should have been able to observe several thresholds. Further, we still see a signal below the threshold, which means there is still at least one channel open at lower energy. Unfortunately, we cannot resolve which of the three possible transitions we have observed, but we can set an interval for the binding energy of this bound mixed state: $0.1584(2) $ eV $ < E_b < 0.2669(2)$ eV. 
\begin{table*}[ht]
    \caption{Potential transitions for the threshold measured at GUNILLA. The first two transitions should be significantly suppressed due to the large $\Delta J$ and detachment of a $d$-electron. Thus, we can conclude that the measured threshold is to the $^3\!F_{5/2}$, $^3\!F_{3/2}$ or $^2\!D_{5/2}$ state in Rh I. This results in a binding energy in the interval  $0.1584(2) $ eV $ < E_b < 0.2669(2)$ eV.}
    \centering
    \begin{tabular}{c | c | c }
        \textbf{Pot. transition from Rh$-$ to Rh I} &\textbf{Rh I energy (eV)} & \textbf{Res. binding energy (eV)}  \\ \hline
    $``^1D_2" \rightarrow (4d^8 5s){}^{3}\!F_{9/2}$  & 0 & $0.5890(2)$ \\
    $``^1D_2"\rightarrow (4d^8 5s){}^{3}\!F_{7/2}$  & 0.189692 & 0.3993(2) \\
    $``^1D_2" \rightarrow (4d^8 5s){}^{3}\!F_{5/2}$  & 0.322115 & 0.2669(2) \\
    $``^1D_2" \rightarrow (4d^8 5s){}^{3}\!F_{3/2}$  & 0.430557  & 0.1584(2) \\
    $``^1D_2"  \rightarrow (4d^9){}^{2}\!D_{5/2}$  & 0.410370  & 0.1786(2) \\
    \end{tabular}
    \label{tab:mixed_transition}
\end{table*}

\begin{figure*}
    \centering
     \begin{subfigure}[b]{0.3\textwidth}
         \centering
         \includegraphics[trim={0cm 0cm 0cm 0cm},clip,width=\textwidth]{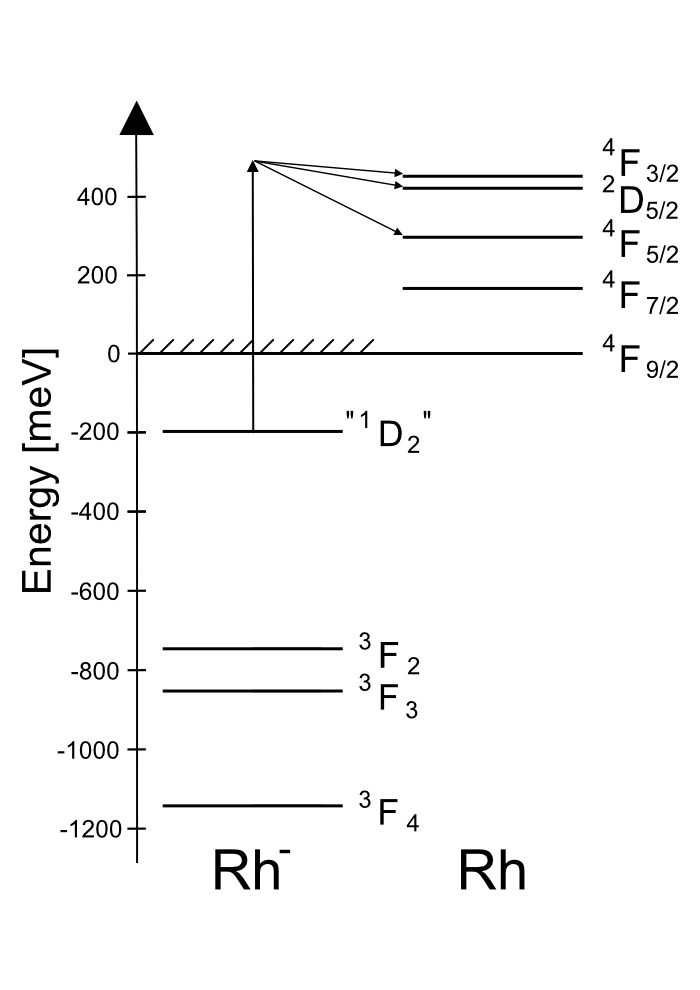}
     \end{subfigure}
        \hfill
    \begin{subfigure}[b]{0.6\textwidth}
        \includegraphics[width = \columnwidth]{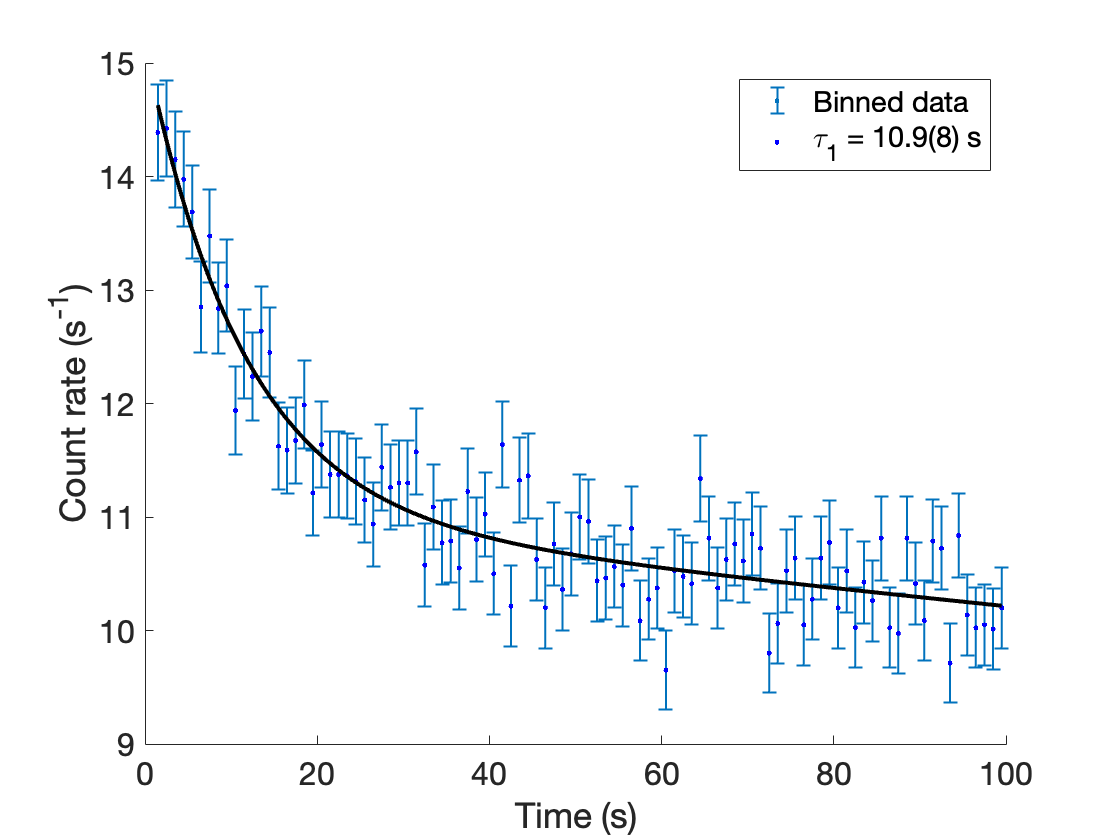}
    \end{subfigure}
        \hfill
    \caption{A typical lifetime measurement of the mixed bound state. The laser wavelength used for this measurement was 1800 nm.}
    \label{fig:topstate1800nm}
\end{figure*}

Using this energy interval in the theoretical lifetime calculations, the lifetime is increased. The VV+CV calculaions, $\tau_{adj}$ ranges from $ 2.7$ s for a binding energy of $0.1584(2) $ eV up to 4.6 s for a binding energy of $0.2669(2) $ eV. 
On the experimental side, the lifetime of this mixed bound state is extracted from a fit to an exponential decay function of the data and is concluded to be 10.9(8) s, as shown in Fig. \ref{fig:topstate1800nm}.

\begin{table*}[]
    \centering
    \caption{Calculated $LS$-composition, excitation energies and radiative lifetimes for the bound states of Rh$^-$. For convenience, the experimental lifetime values are tabulated with both the VV and VV+CV models. }
    \begin{tabular}{c|ccc p{4.4cm} ccccc}
    \toprule
        Model & Label & $J$ & $\pi$ & $LS$-composition & E$_{\rm exc}^{\rm th}$ (eV) &E$_{\rm exc}^{\rm exp}$ (eV) & $\tau_{\rm th}$~(s) & $\tau_{\rm adj}$~(s) & $\tau_{\rm exp}$~(s)\\
     \midrule
        \multirow{7}{*}{VV}& $^3F_4$ & 4 & $+$ & 90$\% \ {}^{3}\!F$ & 0 & 0 & - & - & -\\ 
        & $^3F_3$ &3 & $+$ & 91$\% \ {}^{3}\!F$ & 0.275 & 0.294(8)\footnote{\label{note1}Reference \cite{Feigerle1980BindingIons}.} & 3.5 & 2.9 &  3.2(6)\\ 
        & $^3F_2$ &2 & $+$  & 83$\% \ {}^{3}\!F + 6\% \ {}^{1}\!D$ & 0.417 & 0.418(8) \footref{note1} & 20.4  & 20.2 & 21(4)  \\
        & ``$^{1}\!D_2$" &2 & $+$  & 43$\% \ {}^{1}\!D + 27\% \ {}^{3}\!P + 13\% \ (4d^9 5s)\ {}^{1}\!D + 6\% \ {}^{3}\!F$& 1.037 & $0.8756(2) <E <0.9845(2) $
        & 1.8 &  2.4 - 4.2 & 10.9(8) \\
        & ``$^{3}\!P_2$" &2 & $+$ & $59\% \ {}^{3}\!P + 14\% \ (4d^9 5s)\ {}^{1}\!D + 12\% \ {}^{1}\!D$ & 1.385 & unbound & 4.7 & - & - \\
        \midrule
        \multirow{7}{*}{VV+CV}& $^3F_4$ &4 & $+$ & 90$\% \ {}^{3}\!F$ & 0 & 0 & - & - & -\\ 
        & $^3F_3$ &3 & $+$ & 90$\% \ {}^{3}\!F$ & 0.287 & 0.294(8)  \footref{note1} & 3.1 & 2.9 & 3.2(6)\\ 
        & $^3F_2$ &2 & $+$  & 84$\% \ {}^{3}\!F + 6\% \ ^{1}\!D$ & 0.436 & 0.418(8)  \footref{note1} & 17.4 & 19.8 & 21(4) \\
        & ``$^{1}\!D_2$" &2 & $+$  & 42$\% \ {}^{1}\!D + 35\% \ {}^{3}\!P + 8\% \ (4d^9 5s)\ {}^{1}\!D + 5\% \ {}^{3}\!F$ & 1.107 &  $0.8756(2) <E <0.9845(2) $ 
        & 1.6& 2.7 - 4.6 & 10.9(8)\\
        & ``$^{3}\!P_2$" &2 & $+$  & $53\% \ {}^{3}\!P + 21\% \ {}^{1}\!D + 12\% \ (4d^9 5s)\ {}^{1}\!D $& 1.467 & unbound & 2.3 & - & - \\
        \bottomrule
    \end{tabular}
    \label{tab:LScomp}
\end{table*}

\begin{figure*}
    \centering
     \begin{subfigure}[b]{0.3\textwidth}
         \centering
         \includegraphics[trim={0cm 0cm 0cm 0cm},clip,width=\textwidth]{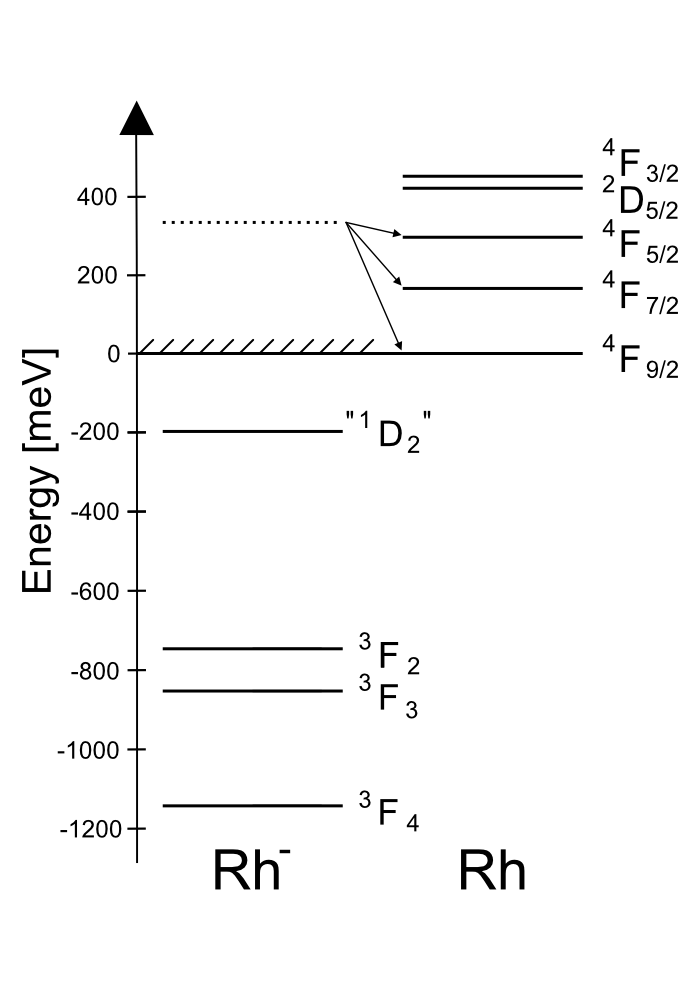}
     \end{subfigure}
        \hfill
    \begin{subfigure}[b]{0.6\textwidth}
        \includegraphics[width = \columnwidth]{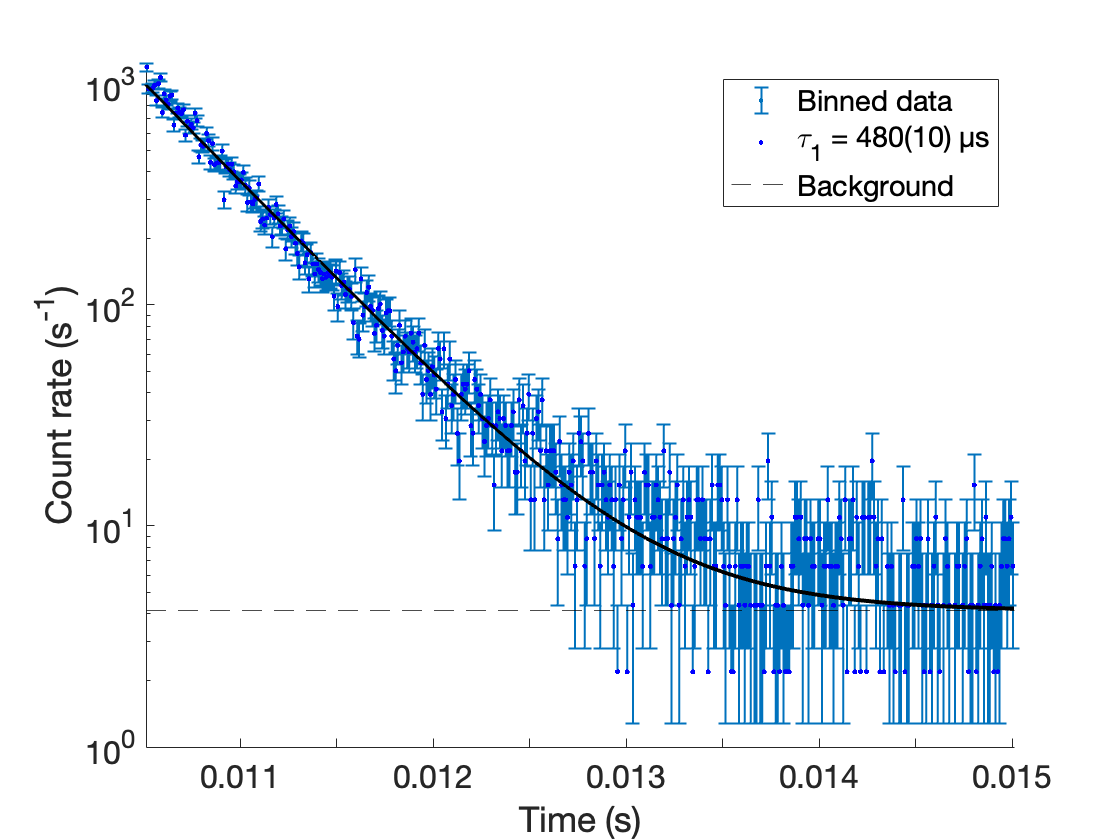}
    \end{subfigure}
        \hfill
    \caption{Typical lifetime measurement of a presumed autodetaching state in Rh$^-$, determined without laser photodetachment.}
    \label{fig:autodetaching}
\end{figure*}

On a very short time scale another signal was observed. This was presumed to be an autodetaching state in Rh$^-$, with a lifetime of 480(10) $\upmu$s as shown in Fig. \ref{fig:autodetaching}.

\section{\label{sec:discussion}Discussion}

There is a good agreement between the theoretically calculated lifetimes of the excited states $^3F_3$ and $^3F_2$ and the experimentally measured values.  For $^3F_3$, the theoretical lifetime adjusted for the measured excitation energy is 2.9(2)~s, which is within the uncertainty of the experimental value 3.2(6)~s.  For $^3F_2$, theory and experiment also agree within uncertainty, with theoretical lifetimes of 20.2(4)~s and 19.8(4)~s from the VV and VV+CV models, respectively, compared to the measured value of 21(4)~s. The challenges are greater for the highest-lying mixed mixed bound state, since the excitation energy of this state is not experimentally known. Although there is a substantial difference between the ab inito lifetime of 1.6(2)~s and the measured value of 10.9(8)~s, the adjusted lifetime approaches the experimental value and the present work still provides valuable information for this previously unobserved excited bound state of Rh$^-$.

In the experimental data, a signal on a short time scale is observed. The energy state responsible for this decay could not be determined. To confirm the existence of this state and not disregard it as an ion beam effect, several measurements were performed at different masses and compared with the data of the autodetaching state in Rh$^-$. The investigations showed that only the Rh$^-$ data seemed to be of an exponential decay type and of a much increased count rate as compared to other measurements; thus the signal in Rh$^-$ could not be disregarded as an ion beam effect. Thus the conclusion is that the observed signal is an autodetaching state. Speculatively, there is a possibility that this autodetaching state could be corresponding to the unbound mixed state, labelled $``^3$P$_2"$ in the theoretical calculations.  

\section{Conclusions}
Photodetachment studies of Rh$^-$ have been performed in order to measure the lifetimes of excited states in the atomic anion. The three bound excited states in the negative ion have been measured to have lifetimes of 3.2(6) s, 21(4) s and 10.9(8) s for the $^3F_3$, $^3F_2$ and the mixed bound states, respectively. The energy level of the mixed state was also determined to be in the interval $0.1584(2) $ eV $ < E_b < 0.2669(2)$ eV. This state had not been previously observed, so the present results provide the first information about this state. In order to experimentally determine the energy state with a higher precision, it would be crucial to identify exactly to which transition the measured threshold energy corresponds. This would require identification of at least one more transition from the mixed state into netural Rh. 

In addition to the lifetimes of the bound excited states in Rh$^-$, a signal on a $\upmu$s time scaled was observed. This was assumed to be an autodetaching state, determined to have a lifetime of 480(10) $\upmu$s, possibly corresponding to the unbound state in the theoretical calculations.

Theoretical calculations of the energies and lifetimes of the excited states were conducted using MCDHF and RCI methods. The calculated lifetimes are in excellent agreement with the measured values for the $^3F_3$ and $^3F_2$ states, demonstrating the high quality of the calculations for predicting M1 decay processes for the fine structure levels in the complex Rh$^-$ ion.  

The calculated lifetime of the $``^1D_2"$ state is very sensitive to changes in the energy due to the very high mixing in this state. The lifetime of this state is mainly decided by the $M1$ transitions to the $^3F_2$ and $^3F_3$ states, as the $E2$ transition is much slower.  Looking at the adjusted lifetime, scaling the energy level to the experimental value in the interval $0.876 $ eV $ < E_b < 0.985$ eV, the lifetime is shifted up to three times the ab initio value, approaching the experimentally measured value. However, the experimental lifetime is still $\sim2-4$ times longer than the theoretical value. This discrepancy shows that our understanding of this highly mixed excited state is still incomplete. 

Combined experimental and theoretical studies such as the present work provide valuable insights into electron correlation in negative ions with general applicability to atomic structure and electric-dipole forbidden radiative transitions in atoms and ions.

\begin{acknowledgments}
This research is supported by the Swedish Research Council Grant No. 2016-03650, 2020-03505 and 2020-05467. This work was supported in part by U.S. NSF Grant No.\ PHY-1707743.
\end{acknowledgments}

\appendix

\bibliographystyle{unsrt}
\bibliography{references,references2}

\end{document}